\documentclass[nofootinbib,pre,showkeys,showpacs]{revtex4-1}
\usepackage{graphicx}
\usepackage{natbib}
\usepackage{bm}
\usepackage{amssymb}
\usepackage{amsmath}
\usepackage{euscript}
\usepackage{mathrsfs}
\usepackage{gensymb}
\usepackage{esint}
\usepackage{slashed}

\begin{document}

\title{Effect of vorticity flip-over on the premixed flame structure: \\ First experimental observation of type I inflection flames}

\author{Hazem El-Rabii}

\affiliation{%
Institute Pprime, CNRS -- UPR 3346, Poitiers, France}%

\author{Kirill A. Kazakov}
\affiliation{%
Moscow State University, Moscow, Russian Federation}

\begin{abstract}
Premixed flames propagating in horizontal tubes are observed to take on shape convex towards the fresh mixture, which is commonly explained as a buoyancy effect. A recent rigorous analysis has shown, on the contrary, that this process is driven by the balance of vorticity generated by a curved flame front with the baroclinic vorticity, and predicted existence of a regime in which the leading edge of the flame front is concave. We report first experimental realization of this regime. Our experiments on ethane and n-butane mixtures with air show that flames with an inflection point on the front are regularly produced in lean mixtures, provided that a sufficiently weak ignition is used. The observed flame shape perfectly agrees with the theoretically predicted.
\end{abstract}
\maketitle

\section{Introduction}

\begin{figure}
\centering
\includegraphics[width=0.7\textwidth]{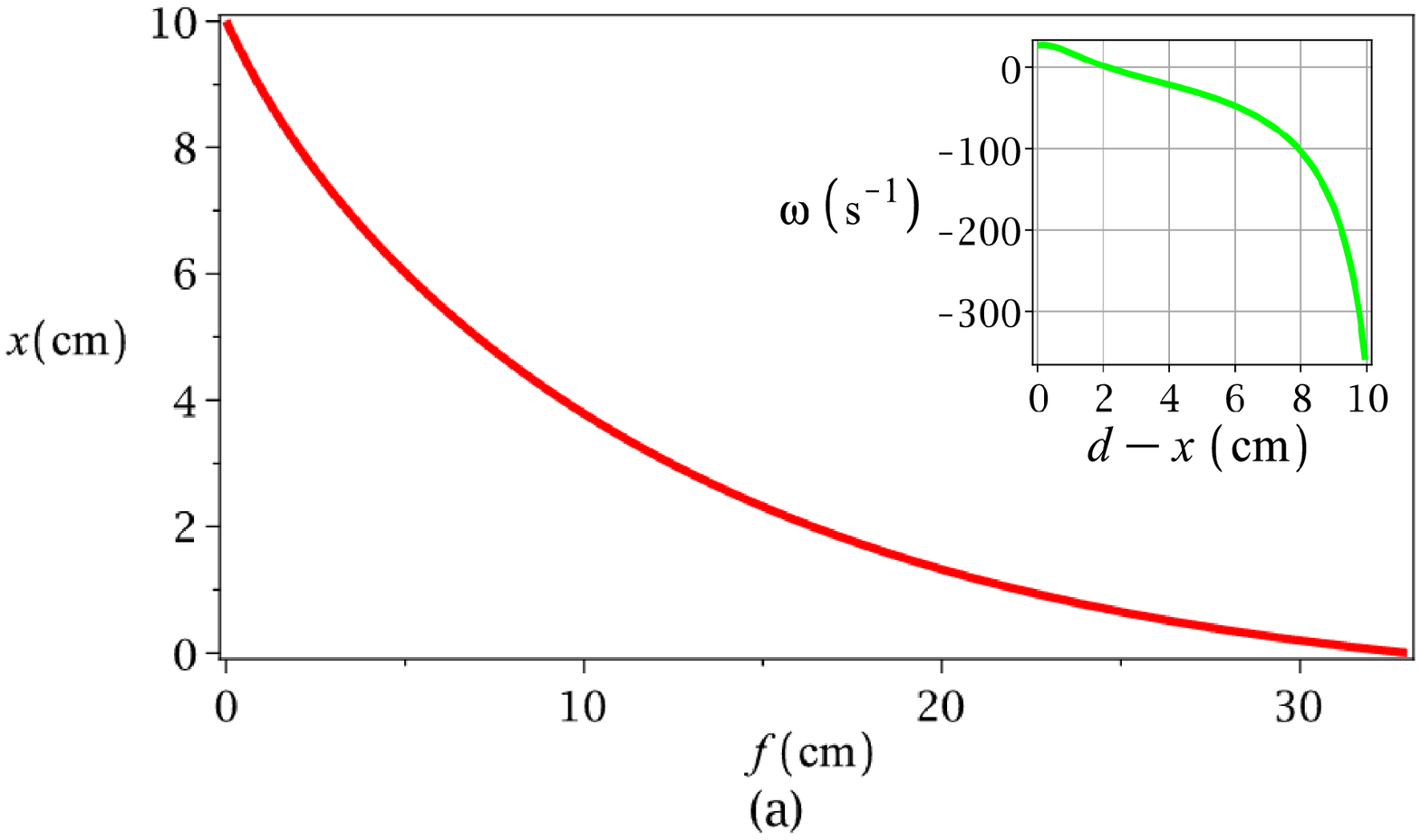}\vspace{2cm}
\includegraphics[width=0.7\textwidth]{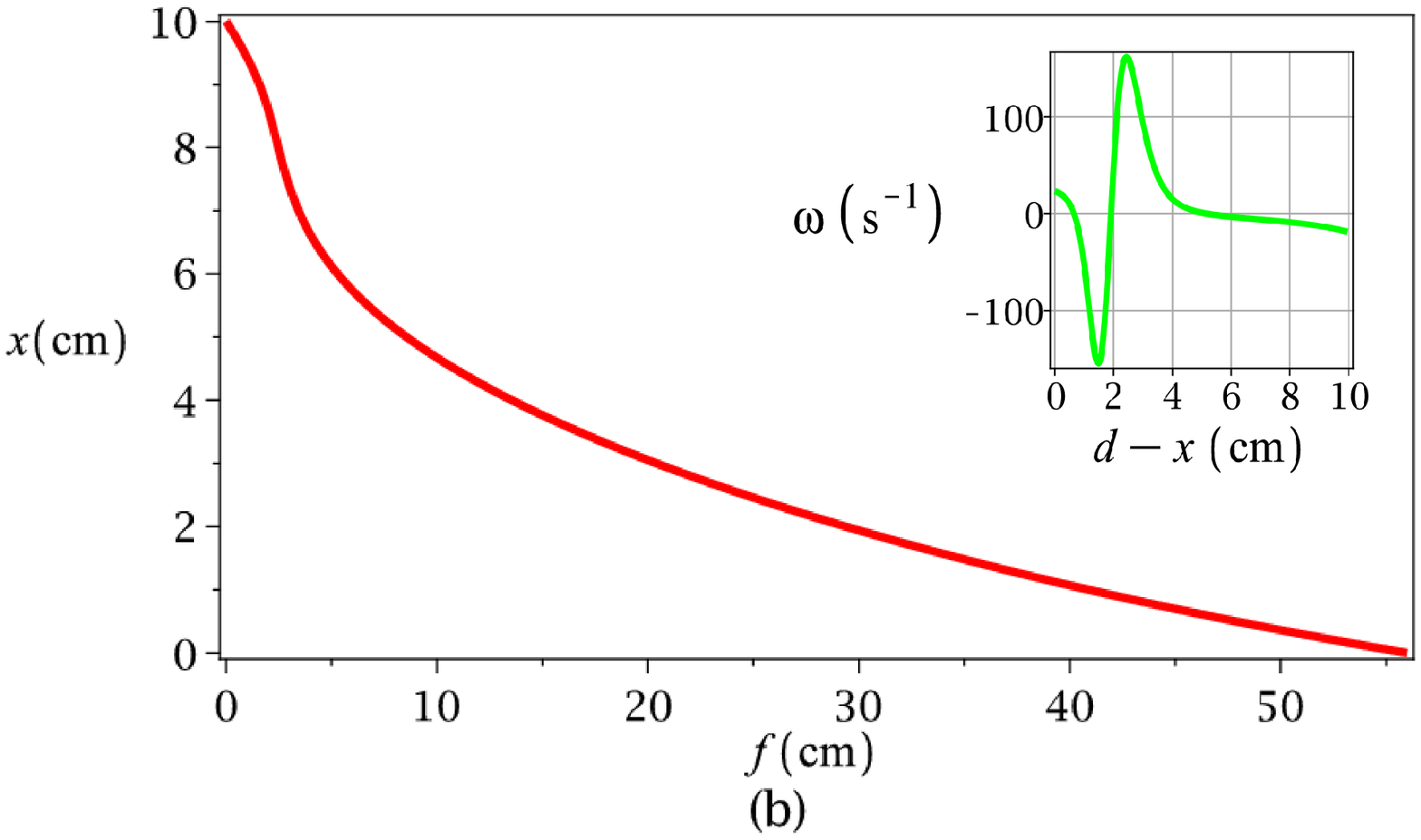}
\caption{(Color online) Front position of conventional flame (a), and type I inflection flame (b) [the inflection point is at $x=8$\,cm]. Insets: burnt gas vorticity at the flame front versus vertical distance from the leading tip. Flames propagate leftwards.}\label{fig1}
\end{figure}

Propagation of a discontinuity surface separating fluids of markedly different densities is an often-encountered physical phenomenon. An air bubble rising in water provides the simplest example, less trivial include laser-driven ablation fronts in plasma \cite{clavin}, condensation discontinuities in supersaturated vapors \cite{landau1}, thermonuclear waves in supernova \cite{oran2005}, but the most prominent instance is the flame propagation in gaseous mixtures \cite{zeldo1985}. Because of the density jump across the surface, its dynamics is affected by gravity, and this influence is stronger on surfaces moving at lower speeds with respect to the fluid. Whenever the mass flux through the surface is nonzero, the same density contrast is also a source of vorticity, so that in all of the above examples except the first, the fluid motion is inherently vortical. Speaking about flames, in the practically most important case of hydrocarbon-air mixtures the process of deflagration is so slow that it is nearly always controlled by terrestrial gravity. At the same time, quantitative description of flame evolution under such conditions is utterly difficult, because one has to deal with extremely nonlinear hydrodynamics of gas flows induced by the flame-gravity interaction, which are characterized by several widely separated length scales. In fact, in the case of flame propagating in a tube, in addition to the tube diameter, $d,$ and the gravitational length $U^2_f/g \ll d$ ($g$ is the gravity acceleration, $U_f$ the planar flame speed relative to fresh gas), there is also the so-called Markstein length $\mathscr{L},$ of the order of the flame front thickness, which characterizes the influence of transport processes inside the front on the flame dynamics. In view of this, it is not surprising that flame evolution in tubes wider than a few centimeters has been primarily an experimental matter \cite{mason1917,coward1932I,searby1992,clanet1998,zipf2014}, and still remains beyond the capabilities of direct numerical methods \cite{liberman2003,michaelis2004,cui2004,kim2006,akkerman2010} as well as of the classical perturbative approach \cite{rakib1988,shtemler1996}. By this reason, an apparent similarity of the first and the last of the above-mentioned  physical systems is often used to simplify modeling of the flame front dynamics \cite{shtemler1996,bychkov1997,bychkov2000}. The underlying hypothesis is that when the normal flame speed relative to the mixture, $U_f,$ is small compared to the bubble speed $\sim\sqrt{gd},$ the flame front shape and its propagation speed can be determined by neglecting the mass flux through the front altogether, that is by taking the limit $U_f/\sqrt{gd}\to 0.$ The gravitational influence on the flame is thus treated as a buoyancy effect. For instance, flames propagating in horizontal tubes are observed to acquire highly elongated shapes, and the textbook explanation is as follows \cite{zeldo1985}: The light products of combustion moving upward tend to occupy the space above heavy fresh gas, which causes the flame front to tilt and spread along the tube, assuming finally a shape convex towards the fresh mixture, Fig.~\ref{fig1}(a). It is only recently that rigorous theoretical assessment of this simplified picture has become possible, with the invention of the on-shell description \cite{kazakov1,jerk}. Namely, the analysis carried out in Ref.~\cite{kazakov2} has shown that neither the limit $U_f/\sqrt{gd}\to 0$ is  well-defined, nor the bubble picture is applicable: flame propagation is actually driven not by buoyancy, but by the baroclinic effect -- gravity-induced vorticity production in the flame front. It is the balance of baroclinic vorticity with that generated by the front curvature which determines the front shape and the associated flow structure. Furthermore, the analysis revealed existence of two essentially different solutions of the on-shell equations. Specifically, it was found that as long as the zero front-thickness is an acceptable approximation, in addition to the observed convex-shaped flames (called type II flames in Ref.~\cite{kazakov2}) there must exist configurations with an inflection point, that is flames having part of the front concave towards the fresh mixture [type I flames, Fig.~\ref{fig1}(b)]. This difference in the front shape is a reflection of the singularity in the vorticity distribution along the front (insets in Fig.~\ref{fig1}): while monotonic in the type II regime, the burnt gas vorticity produced by a type I flame sharply grows and flips over in a vicinity of the inflection point. As this prediction of the theory is so counterintuitive and strikingly distinct from the classical view on flame propagation, its direct experimental verification is indispensable. The present paper reports first realization of type I inflection flames, and the results of comparison with calculations. The experimental setup is described in Sec.~\ref{setup}. Section~\ref{conditions} identifies conditions favorable for the formation of type I inflection flames. Experimental results are presented in Sec.~\ref{exp}, where comparison with the theory is also made. Conclusions are drawn in Sec.~\ref{conclusions}.

\section{Experimental setup}\label{setup}

\begin{figure}
\centering
\includegraphics[width=0.85\textwidth]{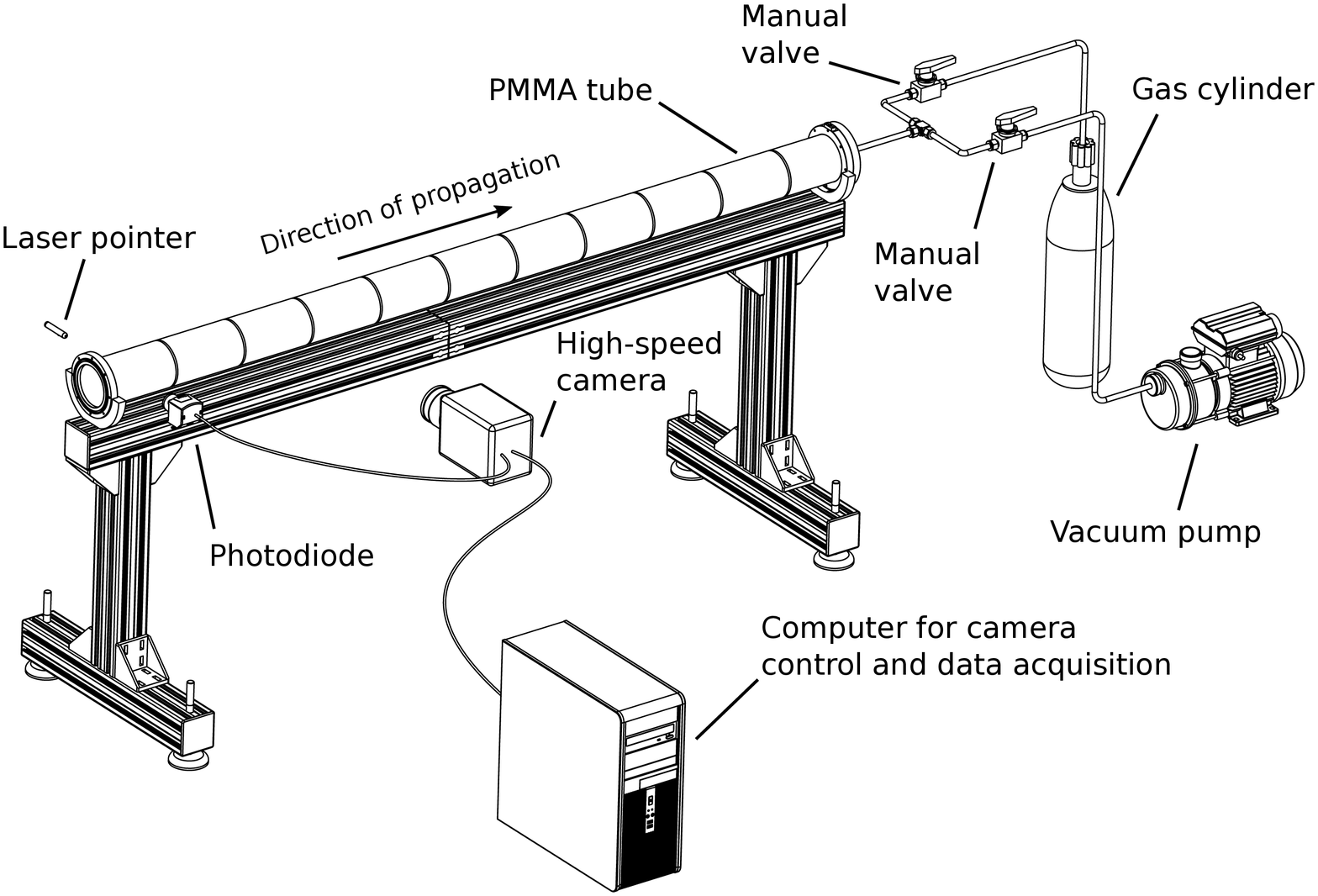}
\caption{Experimental setup.}\label{fig2}
\end{figure}

The flame propagation experiments were carried out in a horizontal circular tube open at the ignition end and closed at the other one. A schematic of the experimental setup used for this study is shown in Fig.~\ref{fig2}. The setup includes a semi-open tube, a high-speed camera, a computer for camera control and data acquisition, a vacuum pump, a metallic cylinder, a laser beam, a photodiode, and a pilot flame.

The tube made of transparent polymethyl methacrylate (PMMA) was 6\,m long with an inner diameter of 10\,cm and a wall thickness of 5\,mm. The fuel-air mixtures investigated were prepared in metallic cylinders by the partial pressure method at a total pressure of 4\,bar, starting from vacuum conditions. Prior to each test, the tube was first closed at both ends, evacuated down to 1\,mbar, and then filled with the desired gas mixture at 1\,bar. Once atmospheric pressure was reached inside the tube, the reactive mixture was left for a few minutes to settle down, whereupon the closing flange was carefully removed. Ignition was achieved with a pilot flame located slightly outside the open end. Flame evolution was recorded with a high-speed camera (Photron RS 3000), its optical axis being directed horizontally perpendicular to the tube axis, at a frame rate of 125\,Hz with an integration time of 3\,ms. The camera was triggered by the photodiode signal produced by the deflection of a laser beam going through the tube, at a 10\,cm distance from its open end. The relative uncertainty of experimental determinations of the flame size and its speed is about $5\%.$

\section{Prerequisites for realization of type I inflection flame}\label{conditions}

\begin{figure}
\centering
\includegraphics[width=0.45\textwidth]{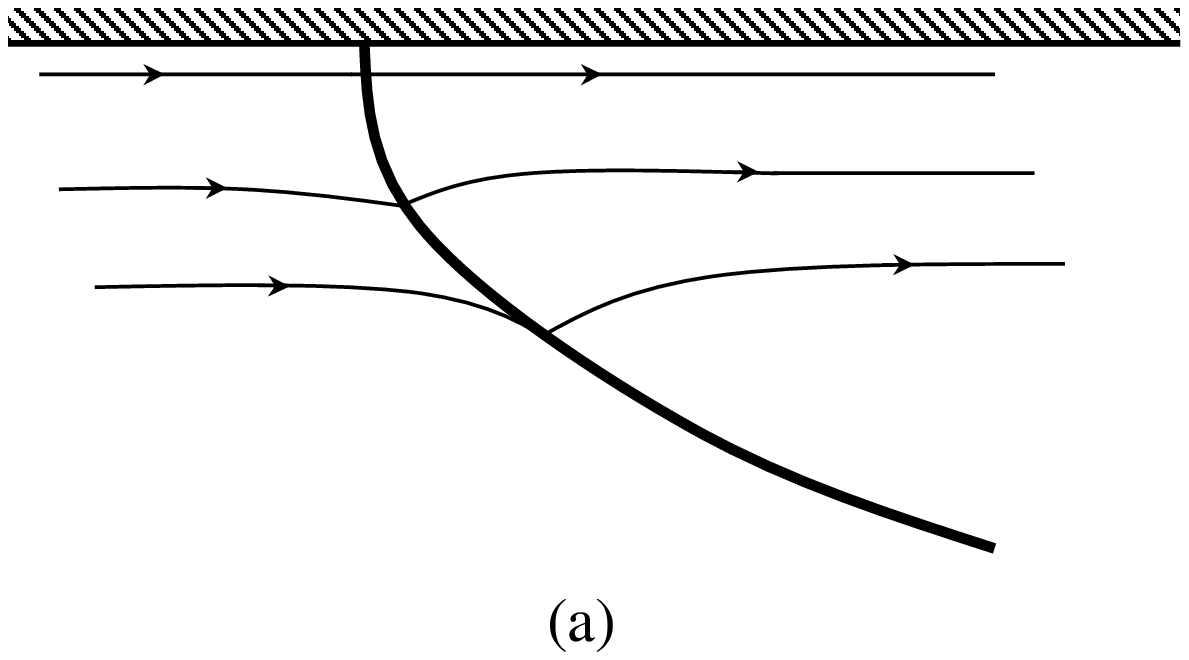}\vspace{2cm}
\includegraphics[width=0.45\textwidth]{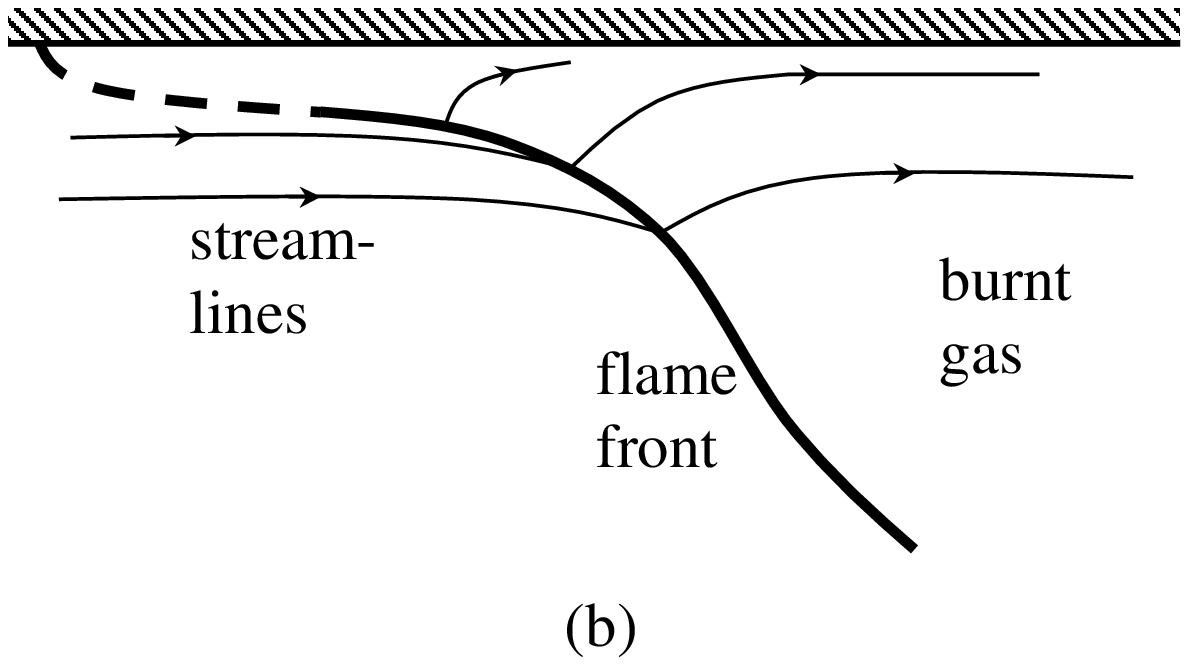}
\caption{Schematics of the gas flow near the upper tube wall. (a) Flame with a convex front as in type II solution shown in Fig.~\ref{fig1}(a). (b) Flame with a front inflection as in type I solution shown in Fig.~\ref{fig1}(b).}\label{fig3}
\end{figure}

The flame profile shown in Fig.~\ref{fig1}(b) was obtained as a solution of the on-shell equations in the zero front-thickness approximation. Real flames have small but finite front thickness, therefore, a successful experimental realization will have to ensure sufficient smallness of the finite front-thickness effects. There is little doubt that the zero front-thickness approximation works well for the most part of the flame in sufficiently wide tubes, because the observed smooth elongated front shapes leave almost no chance for the finite front-thickness effects to show themselves. In fact, the corresponding corrections are described by terms proportional to the small front thickness which involve higher spatial derivatives of the flow fields, so that they become noticeable only for sufficiently corrugated fronts and/or large gradients of the gas velocity. The only region where this approximation is potentially violated is the upper part of the flame including the inflection point. First of all, the front concavity towards the fresh gas makes the burnt gas flow above this point quite distinct from that generated by a convex front. This is illustrated in Fig.~\ref{fig3}. Because of the density drop across the front, the normal gas velocity increases from $U_f$ on the cold side of the front to $\theta U_f$ on its hot side, where $\theta$ is the fresh-to-burnt gas density ratio. As a result, the streamlines sharply bend upwards on crossing the tilted front. After that, they take some space to align with the wall. It is the presence of this transition segment that makes the inflected flame difficult to realize experimentally. Indeed, in contrast to a type II flame where the front convexity allows it to meet the wall at the right angle, Fig.~\ref{fig3}(a), the concavity of inflected type I flame requires an additional front arc (dashed line in Fig.~\ref{fig3}(b)) where the front first turns convex and only then meets the wall. In both type I and type II flames this transition from a finite value of the front slope (corresponding to the left front endpoint in Fig.~\ref{fig1}) to the point at the wall where the front tangent is vertical takes place in a thin layer adjacent to the wall. Flame structure in this layer is controlled by the finite front-thickness effects related to the front curvature, therefore, its thickness is of the order of the cutoff wavelength $\lambda_c.$ The heat outflow from the leading front edge is especially strong in the case of inflected type I flame, because of its peculiar advanced structure described above. As a result, the leading edge temperature, hence its luminosity is significantly reduced.

\begin{figure}
\includegraphics[width=0.5\textwidth]{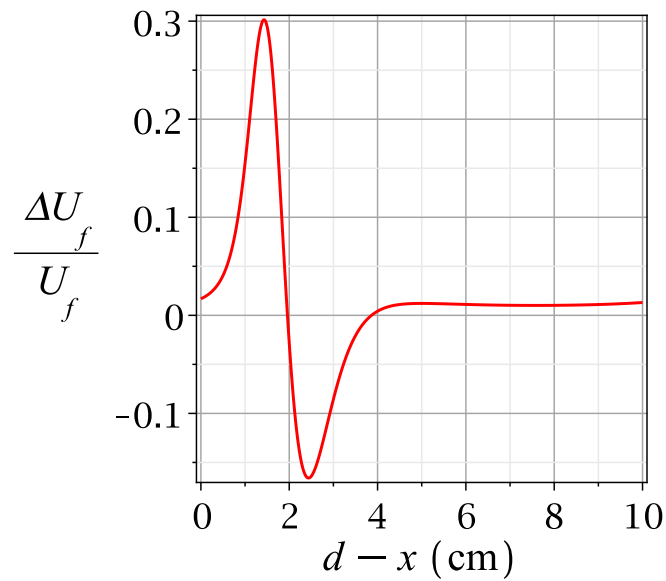}
\caption{(Color online) First-order correction to the normal flame speed, calculated using zeroth order type I solution for a flame with $\theta = 5.64,$ $U_f = 11$\,cm/s. The Markstein length of the flow strain effect $\mathscr{L}_s = -0.45$\,mm \cite{bradley}.}\label{fig4}
\end{figure}

In the case of inflected type I flame, the finite front-thickness effects can  be significant also farther along the front, this time due to the flow strain in a vicinity of the inflection point. As the inset in Fig.~\ref{fig1}(b) shows, type I solutions are characterized by bursts of vorticity near this point; the corresponding fractional correction to the normal front speed, calculated for a lean methane-air flame on a zero front-thickness solution, is plotted in Fig.~\ref{fig4}. It is seen that the flame response on the flow strain in this case is to increase the normal front speed right above the inflection point ($x=8$\,cm), and to decrease it below this point. It follows that this effect, which is due to the high diffusivity of methane, tends to eliminate the front concavity.

\begin{figure}
\includegraphics[width=0.8\textwidth]{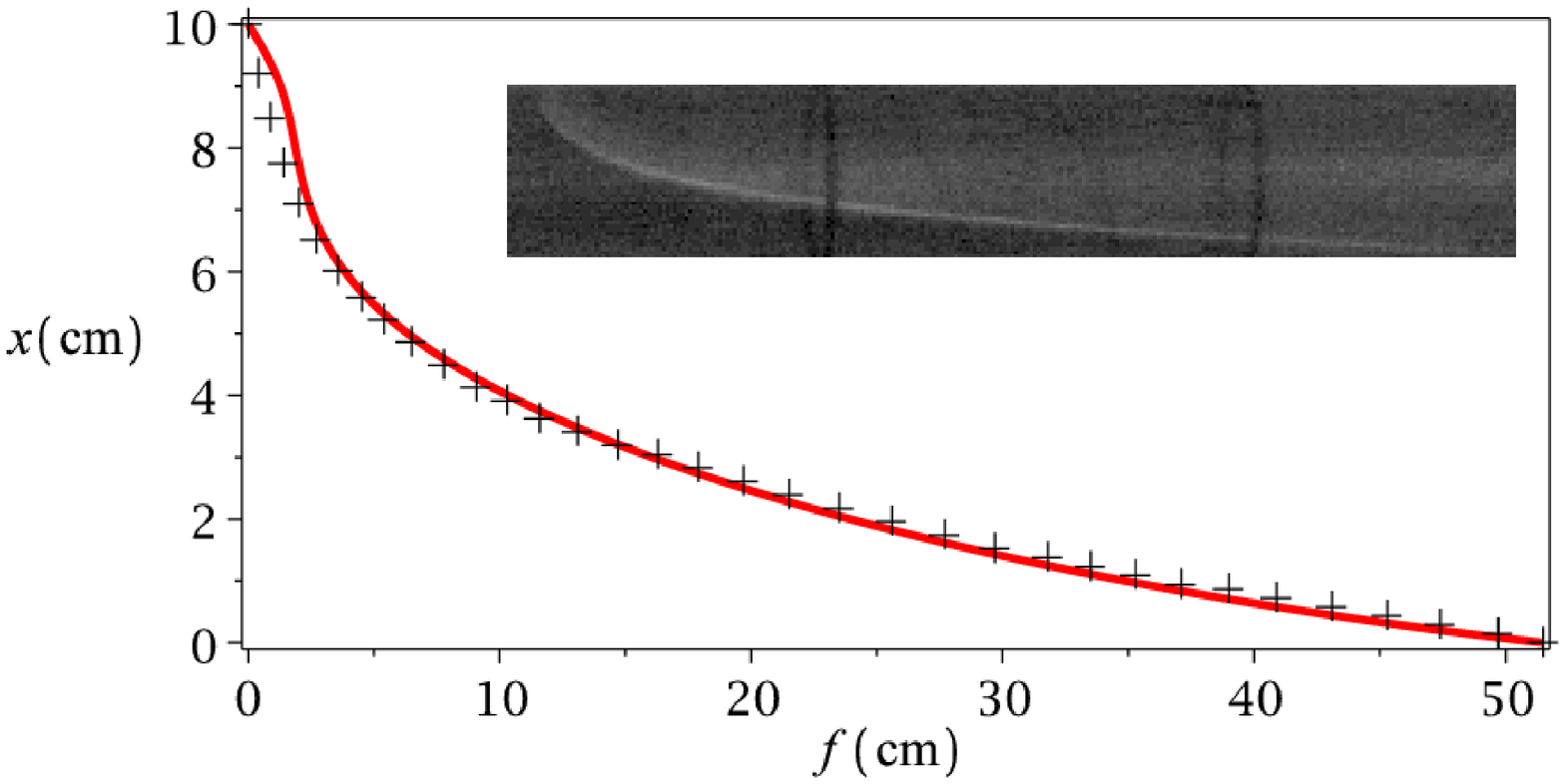}
\caption{(Color online) Front position as given by the type I solution of the on-shell equations for a zero-thickness flame with $\theta = 5.64,$ $U_f = 11$\,cm/s (solid line), and as read off from the photograph of $6\%$ methane-air flame given in the inset (marks).}\label{fig5}
\end{figure}

We thus see that both the front curvature and flow strain effects impede formation of the front inflection. Our observations of methane-air flames show that in most cases they do not develop inflection indeed (in a few instances, however, they did, though we have not been able to identify conditions ensuring reproducible appearance of inflection; it seems that this requires a sufficiently strong flame perturbation). We mention here methane-air flames despite their reluctance to form inflection because it is in this case that sufficiently slow flames were identified in Ref.~\cite{kazakov2} as belonging to type I; this was done indirectly, by analyzing the experimental data \cite{coward1932I} on the flame propagation speed. Therefore, the fact that the observed flame shapes are predominantly convex towards the fresh gas calls for a more detailed experimental verification of the theory. A typical situation is illustrated in Fig.~\ref{fig5} where the observed front shape of $6\%$ methane-air flame is compared with that obtained in Ref.~\cite{kazakov2} as a type I solution of the on-shell equations in the zero front-thickness approximation. The experimental front position is identified hereinafter as the line of maximal brightness on the photograph. The size of the experimental marks (crosses) corresponds to the apparent ``front thickness,'' that is front blurriness resulting from its imperfect flatness; it gives therefore a natural estimate of the experimental error in the front position when compared with the two-dimensional theory \cite{kazakov2}. As expected, the observed front shape noticeably deviates from the calculated only in a small porion of the front surrounding inflection, whereas the rest is well approximated by the type I solution. Since the flame propagation speed is determined by the front length, the speed variation caused by this deviation is well within the calculational error (which is about $10\%$). A consistent account of the finite front-thickness effects on the horizontal flame propagation will be given elsewhere, whereas we turn back to the question of how they can be suppressed.

We note, first of all, that the main factor acting against formation of the front inflection, namely that related to the streamline alignment near the wall, is more pronounced for flames with larger $\theta.$ This is because the jump of the normal gas velocity component at the front is proportional to $\theta - 1 \equiv \alpha,$ hence as $\theta$ increases, more space is required for the gas elements to give away the vertical momentum they received on crossing the front. As this region was treated as negligibly small in the analysis \cite{kazakov2}, applicability of its results can be ensured by increasing the tube diameter $d$ and reducing $\alpha.$ However, in tubes with $d\gtrsim 10$\,cm, flames tend to develop cellular structures complicating the phenomenon to be observed. We therefore take $d=10$\,cm, and turn to flames characterized by lowest $\theta$  -- the so-called cool flames. Cool flames exist at temperatures of a few hundreds degrees Celsius, that is for $\theta$ around $2,$ whereas hot flames have $\theta\gtrsim 5.$ Thus, this gives us a fourfold reduction of $\alpha.$

Second, to reduce as much as possible the finite front-thickness effects related to the flow strain, one has to use mixtures with a sufficiently heavy deficient component. Therefore, we choose lean mixtures of ethane and n-butane with air. These flames have $\lambda_c$ around $3$\,mm, so that condition $d\gg \lambda_c$ is satisfied reasonably well.

\section{Experimental results}\label{exp}

\begin{figure}
\includegraphics[width=1\textwidth]{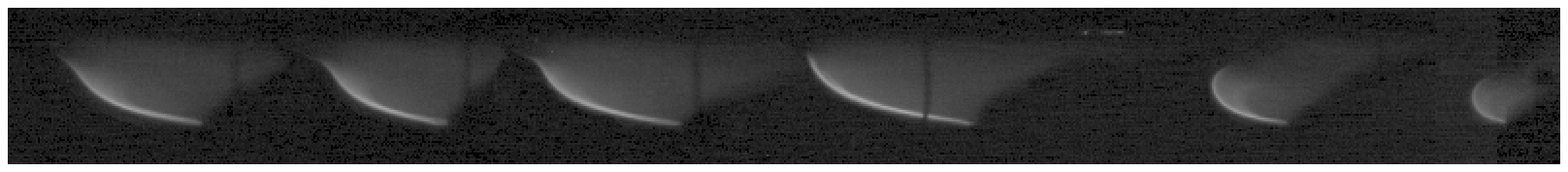}
\caption{Successive images of a cool flame evolution in $2\%$ n-butane-air mixture. Flame propagates from right to left.}\label{fig6}
\end{figure}

\begin{figure}
\centering
\includegraphics[width=0.5\textwidth]{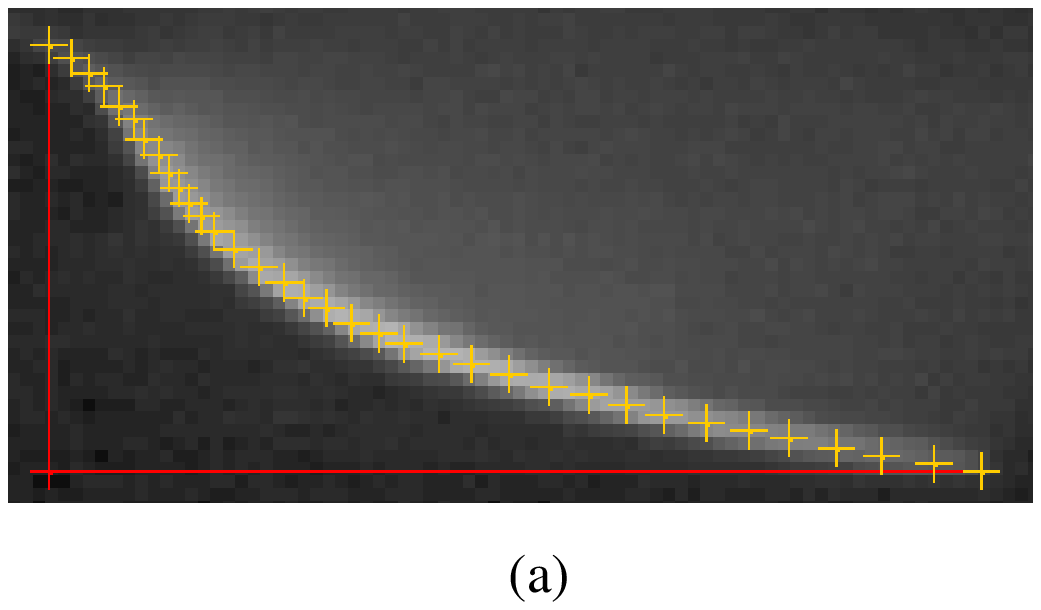}\vspace{2cm}
\includegraphics[width=0.7\textwidth]{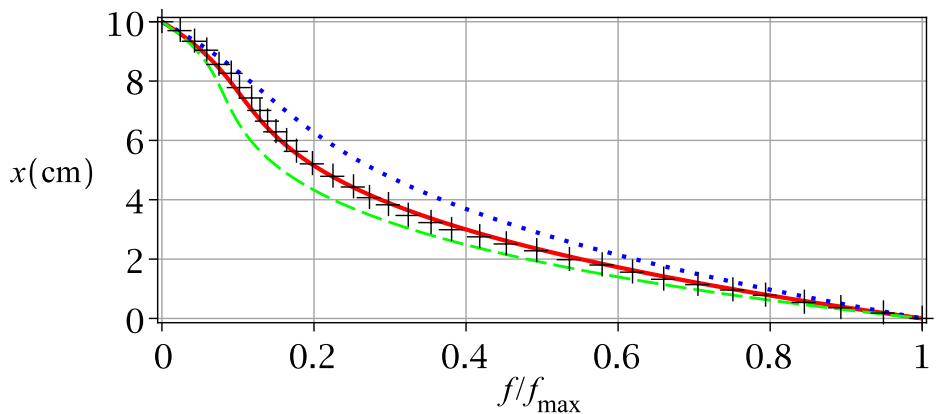}
\caption{(Color online) (a) A digitized image of $2\%$ n-butane-air flame from Fig.~\ref{fig2}. (b) Front shape of steady type I flame with $\theta = 2.2$ (solid), $\theta = 2.5$ (dash), and $\theta = 1.8$ (dot). Flame propagation speed $26.5$\,cm/s. The experimental marks are translated from (a).}\label{fig7}
\end{figure}

As expected of the cool flames, this regime of combustion is obtained when a sufficiently weak energy source is used for ignition. Experimental results are well-reproducible. The sequence of images in Fig.~\ref{fig6} illustrates formation of a type I flame propagating in $2\%$ n-butane-air mixture. It is seen that after a short transient following ignition, a familiar elongated convex flame configuration is formed, which however is also only intermediate, and the flame soon develops an inflection. The inflected flame then propagates steadily over a considerable distance until the slowly growing acoustic instability destroys the steady regime. The observed steady flame shape is compared with the theoretically predicted in Fig.~\ref{fig7}. In order to calculate the front shape, one needs to specify two parameters -- $U_f$ and $\theta.$ Determination of the normal flame speed $U_f$ near the limits of inflammability is not an easy task even for hot flames where the experimental data are largely scattered. Fortunately, the flame shape is practically insensitive to the value of $U_f.$ More precisely, when scaled on its maximal value $f_{\rm max},$ the front position $f$ measured along the tube as a function of the vertical coordinate $x$ is nearly invariant for $U_f$ varying from a few cm/s to $20$\,cm/s. On the other hand, it changes appreciably with $\theta,$ that is with the flame temperature. Cool flames in n-butane-air mixtures are known to belong to the temperature range $200-400\celsius$ \cite{williams1975}, but we were unable to obtain a more accurate value by direct measurements because of large fluctuations in the burnt gas temperature. By this reason, the solid curve in Fig.~\ref{fig6} is drawn as a best fit obtained by varying $\theta$ in the expected range. The value of $\theta$ inferred this way is $2.2,$ which corresponds to $360\celsius$ flame temperature. To illustrate variation of the front position with $\theta,$ two solutions with slightly different $\theta$ and the same flame propagation speed are shown.

\begin{figure}
\includegraphics[width=0.75\textwidth]{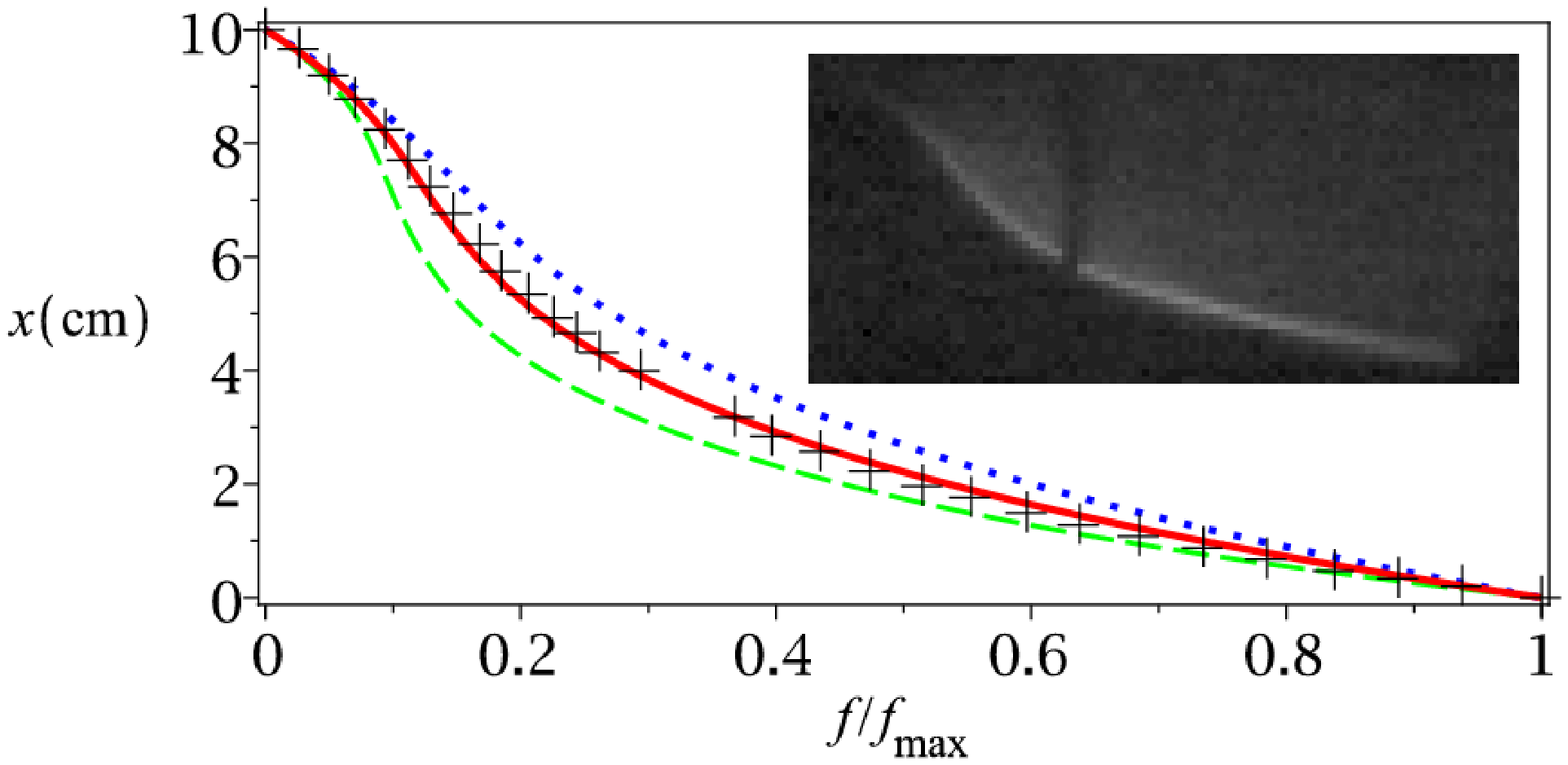}
\caption{(Color online) Front shape of steady flame as observed in $3.6\%$ ethane-air mixture (marks), and calculated for a type I flame with $\theta = 1.8$ (solid), $\theta = 2.0$ (dash), and $\theta = 1.6$ (dot). Inset: the flame photograph. Flame propagation speed $21.5$\,cm/s.}\label{fig8}
\end{figure}

Experiments with ethane-air mixtures gave quite similar results. As an example, Fig.~\ref{fig8} compares the observed front shape of a $3.6\%$ ethane-air steady flame (its image processed as before) and the calculated shapes of type I flames with $\theta = 1.6, 1.8, 2.0.$

\section{Conclusions}\label{conclusions}

Our experiments with lean n-butane-air and ethane-air mixtures have undoubtedly confirmed existence of type I inflection flames. The observed flame front shape is as predicted by the zero front-thickness theory \cite{kazakov2}. Conditions for occurrence of the front inflection are those securing validity of the zero front-thickness approximation for a type I flame -- sufficient thinness of the boundary layer where the flame structure is determined by the front curvature effects, and sufficient smallness of the Markstein length related to the flow strain contribution to the flame stretch, which is significant in a vicinity of the inflection point. The first of these was ensured in our experiments by taking sufficiently wide tube and employing flames with a comparatively small gas expansion. The second is achieved by using mixtures with a sufficiently heavy deficient component.

It is to be stressed that our results not only confirm a particular theoretical prediction, but also disprove the widespread view on the gravity-driven flame propagation. In particular, they illustrate a kind of discontinuity in the dynamics of surfaces with vanishing mass flux. In general, surface dynamics with a small but nonzero mass flux cannot be approximated by that with the strictly zero mass flux, even though this might appear natural on dimensional grounds. In the considered case of premixed flames, the nonexistence of the limit $U_f/\sqrt{gd}\to 0$ makes dynamics of lean flames far more rich than that of bubbles.

\acknowledgments{We gratefully acknowledge the technical assistance of Jean-Carl Rousseau and Vincent Montassier. The study was partially supported by the joint research program CNRS-RFBR (PICS n$^{\circ}$~6147/RFBR 13-02-91054~a).}


\begin{thebibliography}{}

\bibitem{clavin}
P.~Clavin, L.~Masse, and F.~A.~Williams, Combust.~Sci.~Technol.
{\bf 177}, 979 (2005).

\bibitem{landau1}
L.~D.~Landau and E.~M.~Lifschitz, {\it Fluid Mechanics} (Pergamon
Press, New York, 1987).

\bibitem{oran2005}
E.~S.~Oran, Proc.~Combustion~Inst. {\bf 30}, 1823 (2005).

\bibitem{zeldo1985}
Ya.~B.~Zel'dovich, G.~I.~Barenblatt, V.~B.~Librovich, and
G.~M.~Makhviladze, {\it Mathematical Theory of Combustion and Explosions}
(Plenum Press, New York, 1985), pp. 447-450.

\bibitem{mason1917}
W.~Mason and R.~V.~Wheeler, J.~Chem.~Soc., Trans. {\bf 111}, 1044 (1917).

\bibitem{coward1932I}
H.~F.~Coward and F.~J.~Hartwell, J.~Chem.~Soc., 1996 (1932).

\bibitem{searby1992}
G.~Searby, Combust.~Sci.~Technol. {\bf 81}, 221 (1992).

\bibitem{clanet1998}
C.~Clanet and G.~Searby, Phys.~Rev.~Lett. {\bf 80}, 3867 (1998).

\bibitem{zipf2014}
R.~K.~Zipf Jr., V.~N.~Gamezo, K.~M.~Mohamed, E.~S.~Oran, and D.~A.~Kessler, Combust.~Flame {\bf 161}, 2165 (2014).

\bibitem{liberman2003}
M.~A.~Liberman, M.~F.~Ivanov, O.~E.~Peil, D.~M.~Valiev, and
L.~E.~Eriksson, Combust.~Theory~Model. {\bf 7}, 653 (2003).

\bibitem{michaelis2004}
B.~Michaelis and B.~Rogg, J.~Comput.~Phys. {\bf 196}, 417 (2004).

\bibitem{cui2004}
C.~Cui, M.~Matalon, J.~Daou, and J.~Dold, Combust.~Theory~Model. {\bf 8},   41 (2004).

\bibitem{kim2006}
N.~I.~Kim and K.~Maruta, Combust. Flame {\bf 146}, 283 (2006).

\bibitem{akkerman2010}
V.~Akkerman, C.~K.~Law, V.~Bychkov, and L.-E.~Eriksson, Phys.~Fluids {\bf 22}, 053606 (2010).


\bibitem{rakib1988}
Z.~Rakib and G.~Sivashinsky, Combust.~Sci.~Technol. {\bf 59}, 247 (1988).

\bibitem{shtemler1996}
Y.~Shtemler and G.~Sivashinsky, Combust.~Sci.~Technol. {\bf 119}, 35
(1996).

\bibitem{bychkov1997}
V.~V.~Bychkov, Phys.~Rev.~E {\bf 55}, 6898 (1997).

\bibitem{bychkov2000}
V.~V.~Bychkov and M.~A.~Liberman, Phys.~Rep. {\bf 325}, 115 (2000).

\bibitem{kazakov1}
K.~A.~Kazakov, Phys.~Rev.~Lett. {\bf 94}, 094501 (2005); Phys.~Fluids {\bf 17}, 032107 (2005).

\bibitem{jerk}
H.~El-Rabii, G.~Joulin, and K.~A.~Kazakov, Phys.~Rev.~Lett. {\bf 100}, 174501 (2008);
J.~Fluid~Mech. {\bf 608}, 217 (2008); SIAM~J.~Appl.~Math. {\bf 70}, 3287 (2010).

\bibitem{kazakov2}
K.~A.~Kazakov, Phys.~Fluids {\bf 24}, 022108 (2012).

\bibitem{bradley}
D.~Bradley, P.~H.~Gaskell, and X.~J.~Gu, Combust. Flame {\bf 104}, 176 (1996).

\bibitem{williams1975}
F.~W.~Williams, D.~Indritz, and R.~S.~Sheinson, Combust. Sci. Tech. {\bf 11}, 67 (1975).


\end{thebibliography}
\end{document}